\documentclass[aps,prd,twocolumn,notitlepage,10pt,
nofootinbib,nobibnotes,superscriptaddress,preprintnumbers]{revtex4-1}

\usepackage{amsmath}
\usepackage{graphicx,xcolor} 
\usepackage[normalem]{ulem}

\newcommand{\sqrts}{\sqrt{s_{\rm NN}}}
\renewcommand{\d}{\mathrm{d}}

\begin{document}
\title{Friction terms in the multi-fluid description of heavy-ion collisions}
\author{Clemens Werthmann}
\affiliation{Department of Physics and Astronomy, Ghent University, 9000 Ghent, Belgium}
\author{Iurii Karpenko}
\affiliation{Faculty of Nuclear Sciences and Physical Engineering, Czech Technical University in Prague, Břehová 7, 11519 Prague 1, Czech Republic}
\author{Pasi Huovinen}
\affiliation{Incubator of Scientific Excellence—Centre for Simulations of Superdense Fluids, University of Wrocław, Wrocław, 50-204, Poland}
\date{\today} 

\begin{abstract}
    In the multi-fluid description of heavy-ion collisions, the primary scatterings and particle production are described in terms of interaction between fluids, so called friction. These friction terms can be derived from kinetic theory, but they are not unique. We compare different approaches to derive the friction terms, introduce a new ``charge transfer" friction, which allows to move charge to the midrapidity fireball, and implement them in the MUFFIN model. The charge transfer friction is more consistent with the assumption of three fluids clearly separated in momentum space, and allows better comparisons of the experimental data and underlying equation of state. It also leaves room for entropy generation due to dissipation in individual fluids, and we present the first results obtained using viscous multi-fluid dynamics.
\end{abstract}

\maketitle

\section{Introduction}

Heavy ion collisions allow to probe the properties of strongly interacting matter by comparing experimental measurements of the final state particle distributions to theoretical simulations of the collision dynamics. At high collision energies like those at LHC, the theoretical modeling in a multi-stage simulation framework (see, e.g., Refs.~\cite{Putschke:2019yrg,Nijs:2020roc}) has had enormous success due to the fact that the physics at play allows to clearly separate different timescales of the collision: the initial fusion of gluons in the nuclei followed by a gluon dominated phase, then a non-equilibrium evolution of parton-like quasiparticles which quickly approach equilibrium such that hydrodynamics becomes applicable and finally particlization and hadronic transport. In this framework, the role of hydrodynamics is most well-established~\cite{Song:2017wtw,Florkowski:2017olj,Heinz:2024jwu} since most of the observables related to the collective behaviour are built up during the hydrodynamic stage.

At lower energies, such a separation of timescales does not appear. Since the Lorentz $\gamma$ factor is not large, the colliding nuclei have a significant longitudinal size, and the timescale of their interpenetration is comparable to that of the medium evolution. Furthermore, the processes of energy deposition, equilibration and hydrodynamic evolution all happen simultaneously. The new matter constantly coming in from the colliding nuclei causes the system to stay off equilibrium. Additionally, baryons from the colliding nuclei contribute to the plasma evolution but the baryon current decelerates much less than the energy current. Thus, the distribution of particles carrying baryon charge is significantly offset from the overall particle distribution, which is an additional non-equilibrium effect. At collision energies $\sqrts\sim10-100\,$GeV of interest in this work, this baryon transparency is manifested as a double peak structure in the final net baryon distributions. The phenomenon of baryon transparency is already apparent in proton-proton collisions, where the distribution of outgoing baryons is strongly peaked at the edges, while outgoing pions mostly appear at midrapidity.

Hydrodynamic calculations at these energies usually start the hydrodynamic evolution relatively late with an involved and highly parametrized initial
state~\cite{Cimerman:2020iny,Du:2022yok,Jiang:2023fad} or after a preceding transport model description of the primary scatterings~\cite{Auvinen:2013sba, Steinheimer:2014pfa,Karpenko:2015xea,Schafer:2021csj}. Instead of initializing hydrodynamics on a constant (proper) time hypersurface, it is also possible to perform dynamical initialization, where local criteria determine when matter is fed into the hydrodynamic evolution~\cite{ Shen:2017bsr, Akamatsu:2018olk,Shen:2022oyg,Goes-Hirayama:2025nls}.

The multi-fluid model~\cite{Csernai:1982zz,Satarov:1990uc,Mishustin:1991sp,Ivanov:2005yw,Ivanov:2013wha,Ivanov:2013yqa,Ivanov:2013yla,Batyuk:2016qmb,Kozhevnikova:2020bdb,Cimerman:2023hjw,Huovinen:2025lct} aims to describe the low energy nucleus-nucleus collision system by modeling it as multiple interacting fluids with different flow velocities. The incoming nuclei are described as baryon rich ``projectile" and ``target" fluids, which are decelerated by their interaction.  This interaction also produces additional particles in the midrapidity region of the collision, which is modeled as a third ``fireball" fluid.  In momentum space, this description models the system as three distinct thermal distributions that are separated in longitudinal momentum.

The internal dynamics of each fluid is naturally described by hydrodynamics. The key part of a multi-fluid model is the formulation of the interaction between the fluids, called friction, which is not unique. It defines how the fluids affect each other's hydrodynamic variables, such as energy, momentum, and conserved charges. The friction can be derived from the kinetic theory of the underlying microscopic scatterings after choosing a set of rules for which fluid the particles emerging from the scatterings belong to.

Perhaps the simplest friction model was introduced by Csernai \emph{et al.} \cite{Csernai:1982zz}, with an underlying assumption that the energy, momentum, and charges of all particles from the colliding fluids that undergo scattering are immediately assigned to the new fireball fluid. Such simplicity, however, does not allow to describe baryon transparency, i.e. the fact that energy and momentum of the incoming nucleons are stopped stronger than their baryon charges. A somewhat more realistic friction model, that we will refer to as IMS in this paper, was  constructed and amended by Satarov, Mishustin, Russkikh, Ivanov and Toneev in a series of publications \cite{Satarov:1990uc, Mishustin:1991sp, Ivanov:2005yw}. In this model, all baryon charge is kept in the target and projectile fluids, which are decelerated by the friction.

In this study, we introduce a charge transfer friction model based on the ideas we outlined in Ref.~\cite{Huovinen:2025lct}. In this model, outgoing nucleons can be assigned to any of the three fluids depending on their momentum, which more closely adheres to the picture of three separate distributions in momentum space. Furthermore, the charge transfer friction allows the build-up of finite baryon density in the fireball fluid, which makes the model suitable for exploring the equation of state at finite density. We implement the charge transfer friction in the MUFFIN model~\cite{Cimerman:2023hjw}, and study its behaviour via the produced charged hadron and net-proton rapidity distributions. First we briefly compare it to the two prior models, then we explore the effect of varying its free model parameters on the rapidity distributions, and finally we assess how we can best describe experimental data. We notice that the model reproduces the data better if finite shear viscosity is included, and we conclude by showing the results of the first dissipative multi-fluid calculations.

\section{The multi-fluid model}\label{sec:model}

The main idea of a multi-fluid model is to describe the system via three copies of hydrodynamic variables which evolve according to fluid-dynamical evolution equations supplemented with source terms that couple them to each other. Modeling via multiple different fluids makes sense as long as the fluids are clearly separated in momentum space, but in case of degenerate flow velocities, this model artificially increases the number of degrees of freedom. Thus, when the fluids' flow velocities approach each other, they have to be replaced by a single fluid via a procedure called unification. We describe the unification procedure we use in App.~\ref{app:unification}.

In order to allow the fluids to interact, it is not their individual energy and momentum that are conserved, but their sum,
\begin{align}
    \partial_\mu (T^{\mu\nu}_p+T^{\mu\nu}_t+T^{\mu\nu}_f)=0\;,\label{eq:em-conservation}
\end{align}
 where the subscripts $p$, $t$ and $f$ refer to projectile, target and fireball fluids, respectively. This means that the fluids can exchange energy and momentum via so-called friction terms
\begin{align}
    \partial_\mu T^{\mu\nu}_I=F_I^\nu\;,
\end{align}
where Eq.~\eqref{eq:em-conservation} constrains the sum of all friction terms to zero. Analogously, one may also allow the fluids to exchange conserved charges,
\begin{align}
    \partial_\mu N_I^\mu=R_I\;,\quad \sum_I R_I=0\;.
\end{align}

Typically, the friction terms are split into four parts:
\begin{align}
    F_p^\nu=&F_{pt}^{\nu}+F_{pf}^{\nu}\;,\\
    F_t^\nu=&F_{tp}^{\nu}+F_{tf}^{\nu}\;,\\
    F_f^\nu=&-F_{pt}^{\nu}-F_{tp}^{\nu}-F_{pf}^{\nu}-F_{tf}^{\nu}\;.\\
    &\text{(and analogously for the $R_I$)}\nonumber
\end{align}
Of the four terms, $F_{pt}^{\nu}$ and $F_{tp}^{\nu}$, correspond to the transfer from projectile and target to the fireball due to their mutual interaction, so called projectile-target friction. The terms $F_{pf}^{\nu}$ and $F_{tf}^{\nu}$ describe the interchange of energy and momentum between projectile and fireball, and target and fireball, respectively, due to interaction with the fireball fluid, so called fireball friction. Assuming symmetry in the behaviour of projectile and target under friction, the two pairs of friction terms are of the same form and there is no direct transfer between projectile and target due to projectile-target friction. These terms can be derived using kinetic theory, as shown in detail in Apps.~\ref{app:deriving_friction} and~\ref{app:fireball_friction}. However, obtaining any explicit expression for the friction requires choosing which fluids the particles coming out of scatterings are assigned to. One may formulate assignment rules based on both the types and momenta of outgoing particles. Different rules will result in different phenomenology since they approximate the system's dynamics differently.

We compare three different friction models for the projectile-target friction: Csernai-type~\cite{Csernai:1982zz}, Ivanov-Mishustin-Satarov- (IMS-)type~\cite{Satarov:1990uc, Mishustin:1991sp,Ivanov:2005yw} and a charge transfer (CT) friction model, which we have developed. In all three cases, when the friction terms are evaluated, the projectile and target fluids are assumed to consist exclusively of nucleons at zero temperature, while the fireball is made up of pions only. This is a drastic simplification of the system, since we know that heavier resonances and strange particles are formed in the primary collisions too, and since we expect to reach high enough temperatures and densities for the internal degrees of freedom to become partonic. Nevertheless, the cross sections involving heavy resonances and in particular partons are badly or not at all known. For this reason, and also to keep the calculation tractable, we employ the approximative description as nucleonic and pionic degrees of freedom only. To compensate for this approximation, the final friction terms contain scaling terms $\xi_{pt}$ for projectile-target friction and $\xi_f$ for fireball friction, which scale the nucleon densities entering the friction terms.

Note that we use the same fireball friction with all three friction models, and that it is different from the fireball friction employed in the original papers where Csernai- and IMS-type friction were used. In our formulation we assume the fireball fluid to have finite temperature, whereas in IMS the fireball was modeled to be at zero temperature, and in the original Csernai approach there was no fireball friction at all. Furthermore, in the IMS-type fireball friction the outgoing particles were assigned to the projectile and target, whereas we assign them to the fireball. For the details of our fireball friction, see App.~\ref{app:fireball_friction}.

We will now state explicitly how the projectile-target friction takes energy, momentum and conserved charges away from the projectile in the three different models. For now we consider only one charge---baryon charge---but the generalization of the theory to multiple charges is straightforward. As discussed later, all fluids obey the same Equation of State with fixed $n_Q/n_B$ ratio, and therefore baryon transfer implies electric charge transfer as well. The effect of friction on the target is obtained by interchanging the variables related to the projectile and target ($t\leftrightarrow p$). In the Csernai model, all particles that scatter will be assigned to the fireball. This makes it easy to compute friction as the integral over all conserved quantities carried by projectile and target particles undergoing scattering:
\begin{align}
    R_{B,pt,\rm Csernai} =&-\xi_{pt}^2n_p n_t V_{\rm rel}^{pt}\sigma_{NN\to X}(s_{pt})\;,\\
    F_{pt,\rm Csernai}^{\nu}=&-\xi_{pt}^2m_Nn_pn_tu^\nu_pV_{\rm rel}^{pt}\sigma_{NN\to X}(s_{pt}).
\end{align}
Here, $n_I$ is the nucleon density of fluid $I$ and $u_I^\nu$ its flow velocity, $m_N$ is the nucleon mass, $\sigma_{NN\to X}$ is the total nucleon-nucleon cross-section, $s_{pt}=m_N^2(1+2u_{p,\mu}u_t^\mu)$ and $V_{\rm rel}^{pt}=\sqrt{(u_{p,\mu}u_t^\mu)^2-1}$.

In IMS friction, all created pions go to the fireball, but all nucleons stay in projectile and target. Which one they will be assigned to depends on their rapidity. This choice results in
\begin{align}
    R_{B,pt,\rm IMS} =&0\;,\\
    F_{pt,\rm IMS}^{\nu}=&-\xi_{pt}^2m_Nn_pn_tV_{\rm rel}^{pt}\\
    &\times\frac{1}{2}\big[(u_p^\nu-u_t^\nu)\sigma_P(s_{pt})+(u_p^\nu+u_t^\nu)\sigma_E(s_{pt})\big]\;,\nonumber
\end{align}
where
\begin{align}
    \sigma_E(s)=&\int_{w_p}\d\Gamma \frac{\d\sigma_{NN\to NX}(s)}{\d\Gamma}\left(1-\frac{E}{E_0}\right)\;,\\
    \sigma_P(s)=&\int_{w_p}\d\Gamma \frac{\d\sigma_{NN\to NX}(s)}{\d\Gamma}\left(1-\frac{p_L}{p_0}\right)\;,
\end{align}
with $E_0$ and $E$ being the in- and outgoing nucleon energies, $p_0$ being the ingoing nucleon momentum, $p_L$ being the outgoing nucleon longitudinal momentum and integration being performed over all transverse and in the CMS frame positive longitudinal momenta, $w_p=\{p_L>0\}$, with the measure $\d\Gamma={\d^3p}/{[(2\pi)^3p^0]}$.

The new charge transfer friction can be understood as a generalization of the IMS friction, where outgoing nucleons at small rapidities are assigned to the fireball. This choice is more faithful to the justification of the multi-fluid model having fluids separated in momentum space. Also, it is helpful for probing the equation of state, as it allows the fireball to acquire a finite net-baryon density. Specifically, outgoing nucleons with rapidities below a threshold rapidity $\beta\cdot y_{\rm in}$ are assigned to the fireball fluid and nucleons at higher rapidities to the original projectile or target fluids, where $y_{\rm in}$ is the rapidity of ingoing nucleons in the CMS frame and $\beta$ is a new parameter of the model. Therefore, the treatment of nucleons in this model is interpolated between Csernai-type friction at $\beta=1$ and IMS-type friction at $\beta=0$. The former limit faithfully reproduces Csernai friction, but the latter is not fully identical to IMS friction. This is because the correlation of the outgoing nucleon's energy with its rapidity is not known, so the charge transfer model also introduces a second parameter $\alpha$ that quantifies what fraction of the energy coming from the decrease in nucleon rapidity is assigned to the fireball. In other words, $\alpha$ controls how much energy is used to create new particles in nucleon-nucleon scattering. In the IMS case, all nucleons stay in projectile and target, so knowledge of this correlation is not needed. The energy transfer is simply fixed by the mean energy of all outgoing pions, which is experimentally known as a function of $\sqrt{s}$. The CT friction terms are given by
\begin{align}
    R_{pt,\rm CT} =&-\xi_{pt}^2n_pn_tV_{\rm rel}^{pt}\bar{\sigma}_R(s_{pt})\;,\\
    F_{pt,\rm CT}^{\nu}=&-\xi_{pt}^2m_Nn_pn_tV_{\rm rel}^{pt}\Big\{\frac{1}{2}\big[(u_p^\nu-u_t^\nu)\bar{\sigma}_P(s_{pt})\\
    &+(u_p^\nu+u_t^\nu)\bar{\sigma}_E(s_{pt})\big]+u_p^\nu\bar{\sigma}_R(s_{pt})\Big\}\;,\nonumber
\end{align}
where
\begin{align}
    \bar{\sigma}_E(s)=&\int_{\bar{w}_p}\d\Gamma \frac{\d\sigma_{NN\to NX}(s)}{\d\Gamma}\left(1-\frac{E}{E_0}\right)\;,\\
    \bar{\sigma}_P(s)=&\int_{\bar{w}_p}\d\Gamma \frac{\d\sigma_{NN\to NX}(s)}{\d\Gamma}\left(1-\frac{p_L}{p_0}\right)\;,\\
    \bar{\sigma}_R(s)=&\int_{\bar{v}_p}\d\Gamma \frac{\d\sigma_{NN\to NX}(s)}{\d\Gamma}\;,
\end{align}
with $\bar{w}_p=\{y>\beta y_p\}$ and $\bar{v}_p=\{0<y<\beta y_p\}$.

 We solve the evolution equations using MUFFIN~\cite{Cimerman:2023hjw}, our multi-fluid solver based on the vHLLE code~\cite{Karpenko:2013wva}. To describe the hadronic stage and calculate final hadron distributions, we switch to the SMASH hadron cascade~\cite{SMASH:2016zqf}. In the calculations presented here, we use event averaged initial states. The setup of the two colliding nuclei in the initial state and details of particlization are explained in Ref.~\cite{Cimerman:2023hjw}. As in Ref.~\cite{Cimerman:2023hjw}, we use the chiral model based Equation of State (EoS)~\cite{Steinheimer:2010ib} constrained to zero strangeness density, $n_S=0$, and to electric charge density $n_Q=0.4\,n_B$. Thus the EoS depends only on energy and net-baryon densities. At particlization we employ the same constraints. To evaluate the friction terms, we approximate the nucleon density by the net-baryon density of each fluid, and as the temperature of the pions in the fireball, we use the temperature of the fireball fluid. Likewise, the temperature required for the unification procedure (see App.~\ref{app:unification}) is the temperature of the projectile and target fluids.

\begin{figure*}[ht]
 \centering
 \includegraphics[width=0.49\textwidth]{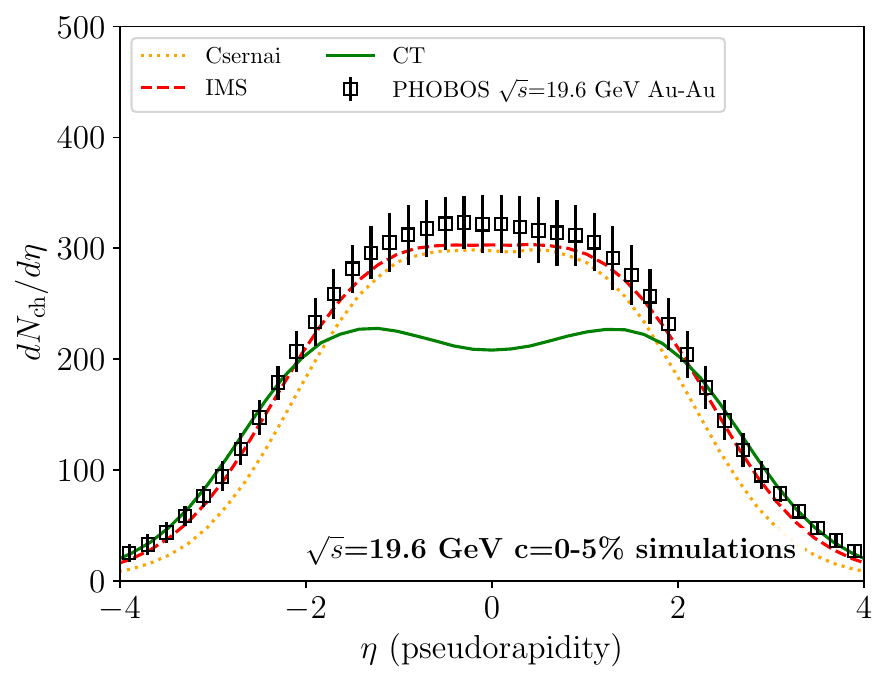}
 \includegraphics[width=0.49\textwidth]{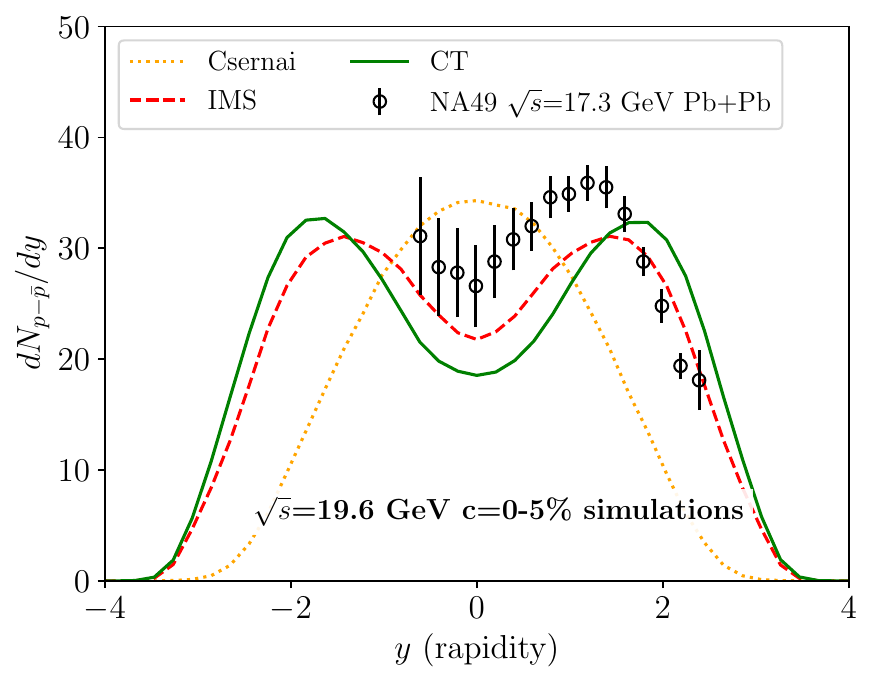}
 \caption{Pseudorapidity distributions of all charged hadrons (left) and rapidity distribution of net protons (right) with the Csernai, IMS and CT friction models  using parameter values $\xi_{\rm pt}=1.0$, $\xi_{\rm f}=0.1$ for all models and $\alpha = 0.7$, $\beta = 0.1$ for CT friction. The data are by the PHOBOS~\cite{PHOBOS:2010eyu} and the NA49~\cite{NA49:1998gaz} collaborations.}\label{fig:FMs}
\end{figure*}

In all works on multi-fluid dynamics that have been published so far, the internal evolution of the fluids has been described by ideal hydrodynamics. When exploring the effects of the model parameters, we keep this tradition. However, as described later, the charge transfer friction model can reproduce experimental data better if shear viscosity of the fluids is included. At this stage, we include only the shear stress tensor in the internal evolution of the fluids. In the current implementation, friction terms do not directly affect shear, but do affect the Navier-Stokes value entering its evolution equation via the changes in the velocity fields. Additionally, the code will cap the size of the shear tensor whenever it gets too large relative to energy and momentum.

With the inclusion of the shear stress, we need to model the shear viscosity $\eta$. We employ a recently introduced parametrisation from Ref. \cite{Jahan:2024wpj}:
\begin{multline}
\tilde{\eta}(\mu_B) =\\ \begin{cases}
    \eta_0 + (\eta_2 - \eta_0)\frac{\mu_B \text{[GeV]}}{0.2}, & 0 < \mu_B \leq 0.2 \text{\,GeV}, \\
    \eta_2 + (\eta_4 - \eta_2)\frac{(\mu_B \text{[GeV]} - 0.2)}{0.2}, & 0.2 < \mu_B < 0.4 \text{\,GeV}, \\
    \eta_4, & \mu_B \geq 0.4 \text{\,GeV},
\end{cases}\label{eq:etaS}
\end{multline}
where $\tilde{\eta}=\eta\cdot T/(\varepsilon+p)$ is the ratio of shear viscosity to enthalpy density, which is set to depend only on the baryon chemical potential. We employ the parameter values $\eta_0=0.045$, $\eta_2=0.28$ and $\eta_4=0.287$ obtained in the Bayesian analysis of~\cite{Jahan:2024wpj}. Note that in this paper, we use the same symbol $\eta$ for both shear viscosity and pseudorapidity. To avoid confusion, we will spell out the name of the respective quantity when it appears.

For the non-equilibrium correction $\delta f$ to the hadron distribution function $f(x,p)$ at particlization, we use the well-known Grad’s 14-moment ansatz for a single component system, and assume that the correction is the same for all hadron species:
\begin{equation}
\delta f_i = f_{{\rm eq}, i} (1 \pm f_{{\rm eq}, i}) \left[ \frac{p^\mu p^\nu \pi_{\mu\nu}}{2T^2(\varepsilon + p)} \right]\ ,
\end{equation}
where $f_{{\rm eq}, i}$ is the equilibrium distribution function for the given hadron species $i$.

\section{Friction model benchmarks}

\begin{figure*}[ht]
 \centering
 \includegraphics[width=0.49\textwidth]{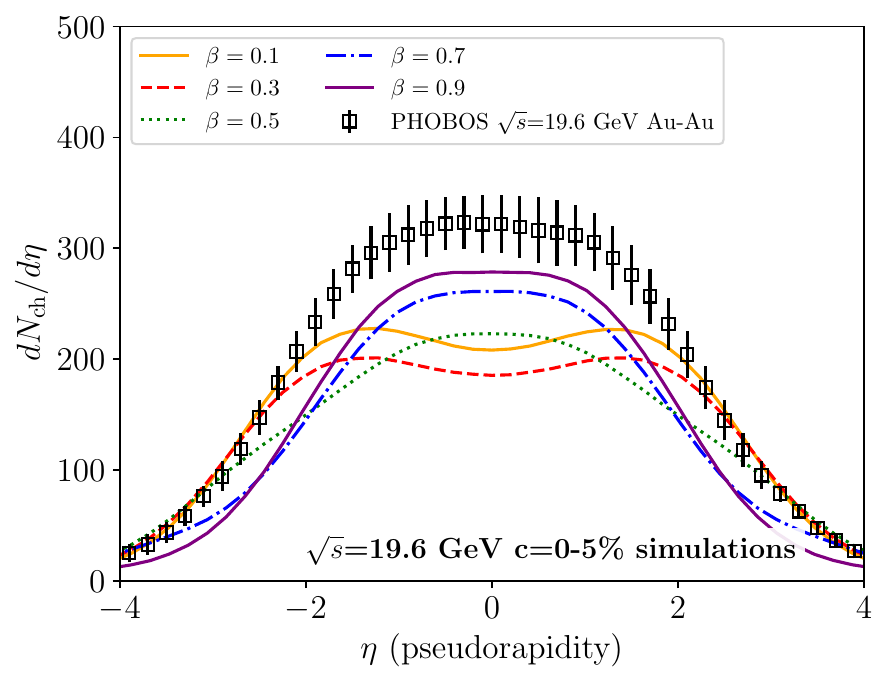}
 \includegraphics[width=0.49\textwidth]{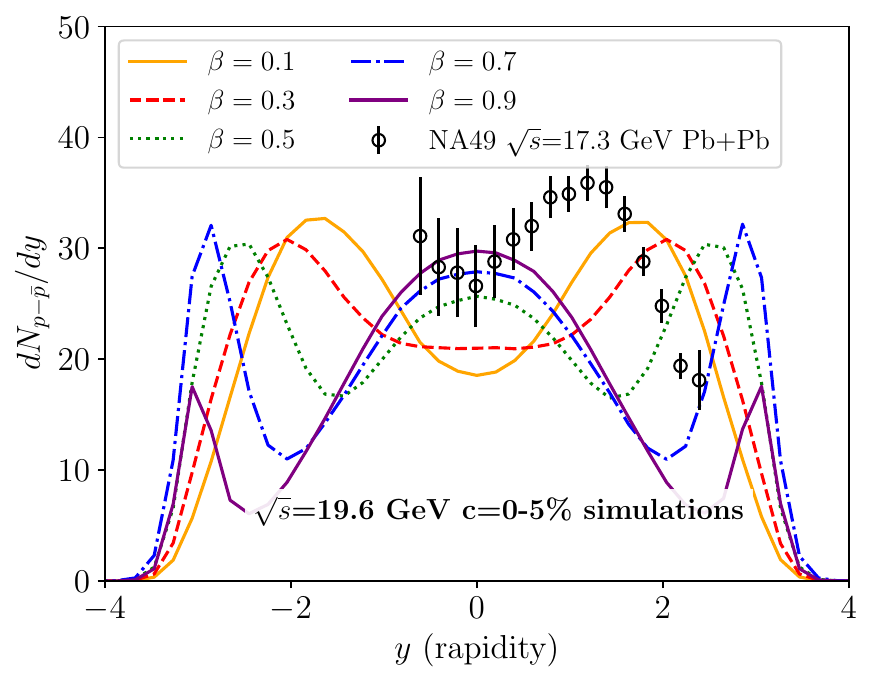}\\
 \includegraphics[width=0.49\textwidth]{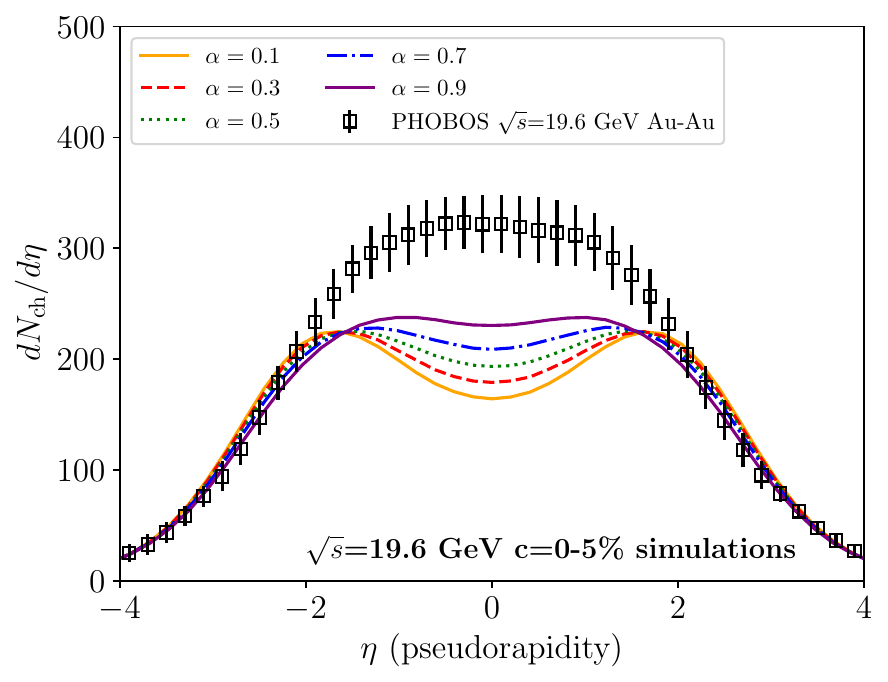}
 \includegraphics[width=0.49\textwidth]{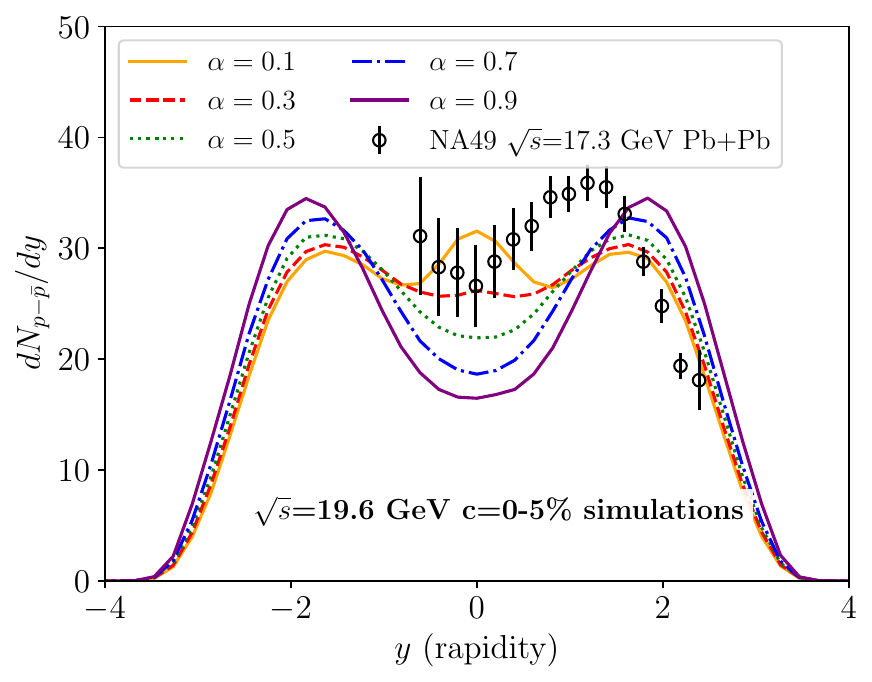}
 \caption{Pseudorapidity distributions of all charged hadrons (left) and rapidity distribution of net protons (right) with the CT friction model. Scans over the $\beta$ and $\alpha$ parameters are conducted using $\xi_{\rm pt}=1.0$, $\xi_{\rm f}=0.1$, $\alpha=0.7$ (for top row) and $\beta=0.1$ (for bottom row). The data are by the PHOBOS~\cite{PHOBOS:2010eyu} and the NA49~\cite{NA49:1998gaz} collaborations.}\label{fig:alpha-beta-scan}
\end{figure*}

Since the different friction models lead to different longitudinal dynamics, we describe their effects on two final-state hadronic observables: pseudorapidity distribution of all charged hadrons and rapidity distribution of net protons. The former is sensitive to the entropy production and deposition throughout the collision, whereas the latter characterises baryon stopping. We chose the 0-5\% centrality class of collisions at an energy of $\sqrts=19.6\,$GeV to benchmark the friction models. At this value of the collision energy the double peak structure due to baryon transparency is already visible in the net-proton distribution. Another advantage of this choice of collision energy is that both the pseudorapidity distribution of charged hadrons (from the PHOBOS experiment at $19.6\,$GeV \cite{PHOBOS:2010eyu}) and the rapidity distribution of net protons (from the NA49 experiment at $17.3\,$GeV \cite{NA49:1998gaz}) are available at very close collision energies.

The results from different friction models are shown in Fig.~\ref{fig:FMs}. The Csernai friction model produces a reasonable shape of the charged hadron pseudorapidity distribution $dN_{\rm ch}/d\eta$, but fails to reproduce the experimental rapidity distribution of net protons. Because of its basic assumption that the energy, momentum and charges of all particles undergoing scatterings are immediately assigned to the fireball fluid, the transport of charges behaves exactly like the transport of energy. Therefore, there is strong baryon stopping at all collision energies. IMS-type friction improves on the prediction of the net-proton distribution as it keeps all of baryon charge in the target and projectile fluids, which are only slowly decelerated by the friction. This allows to produce a double peak in the distribution as seen in Fig.~\ref{fig:FMs}.

\begin{figure*}[ht]
 \centering
 \includegraphics[width=0.49\textwidth]{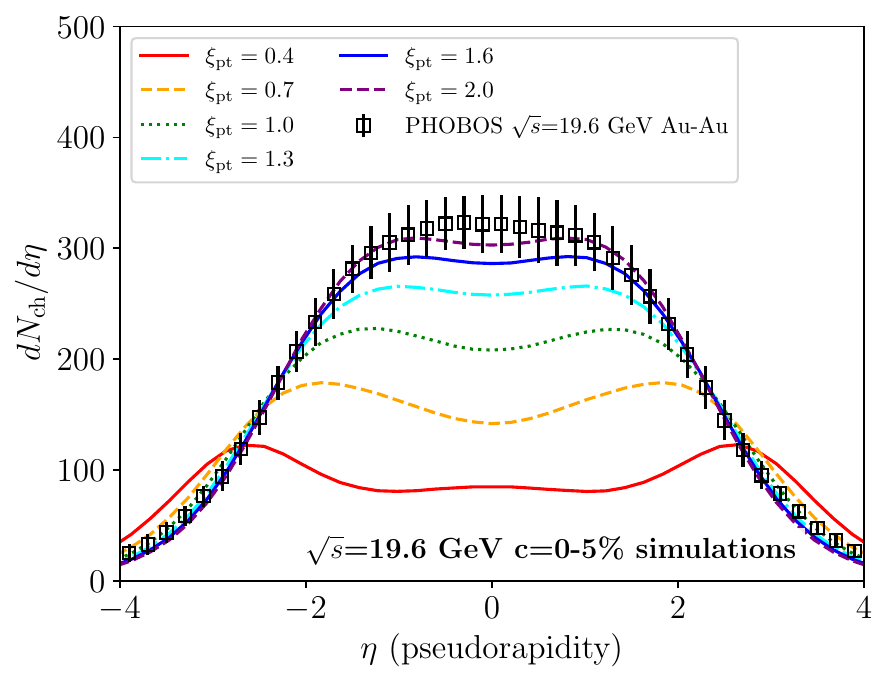}
 \includegraphics[width=0.49\textwidth]{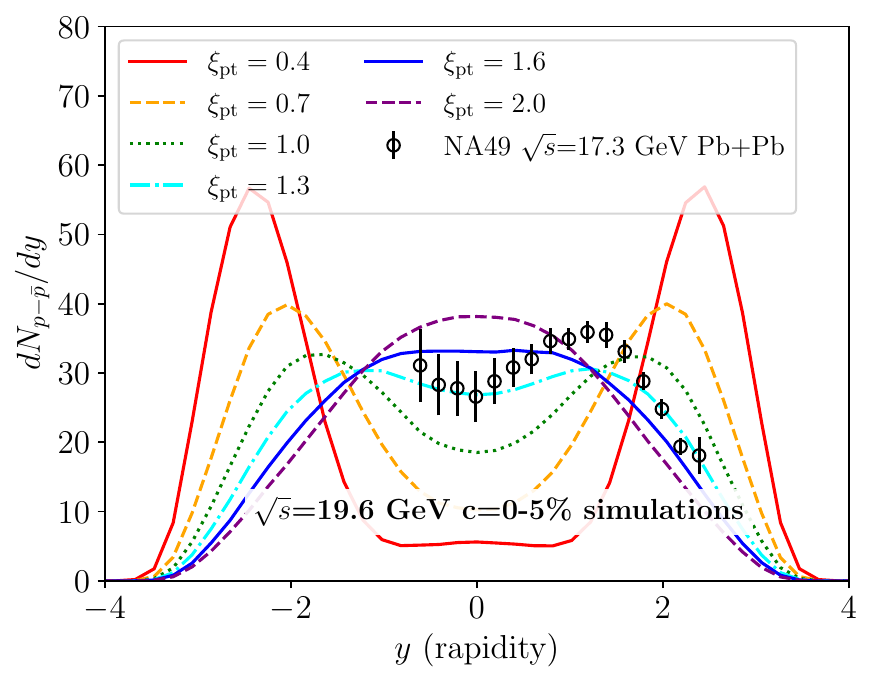}\\
 \includegraphics[width=0.49\textwidth]{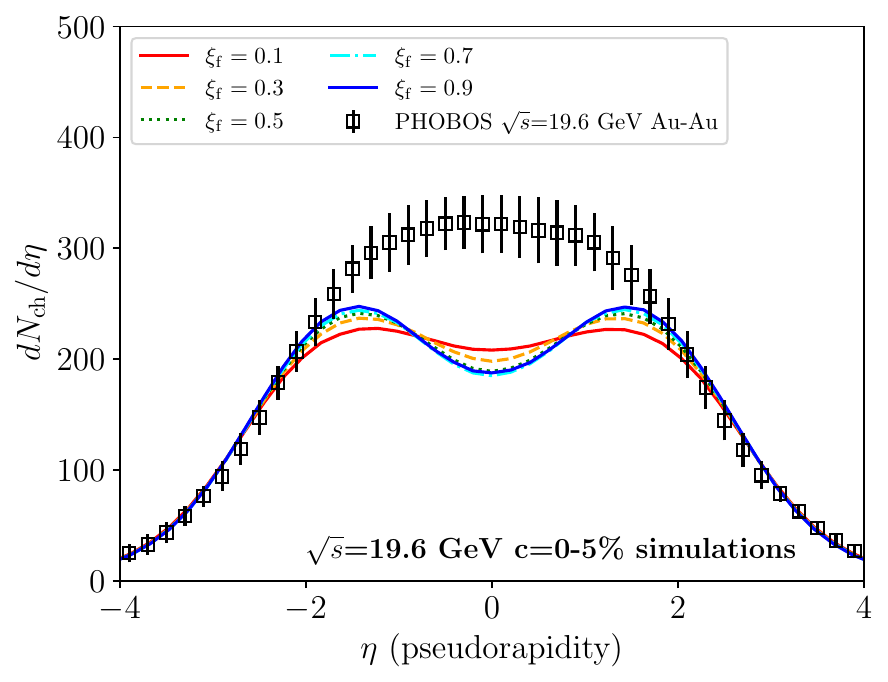}
 \includegraphics[width=0.49\textwidth]{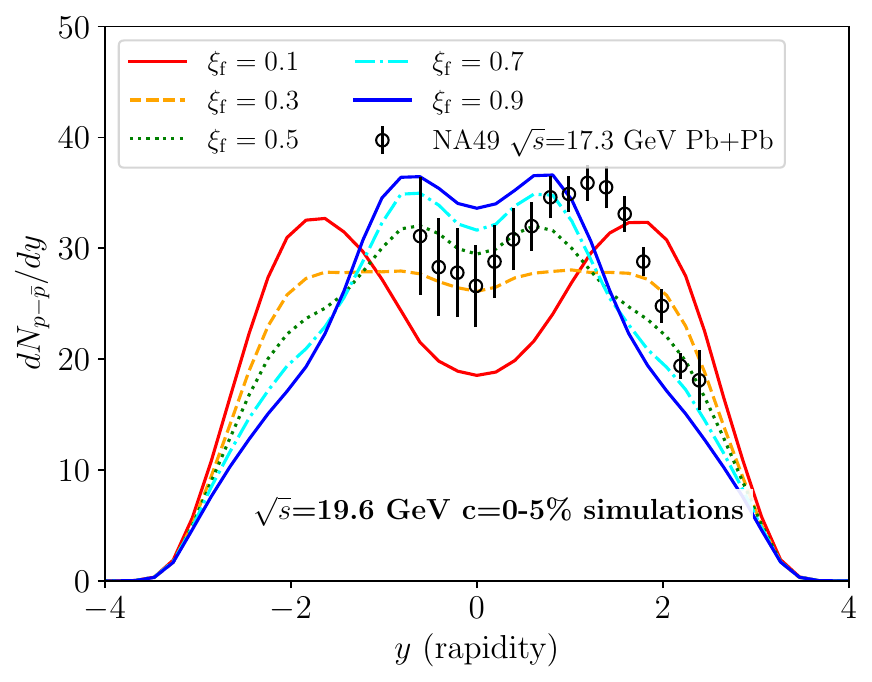}
 \caption{Same quantities as in Fig.~\ref{fig:alpha-beta-scan}, but simulated with different projectile-target friction scaling ($\xi_{\rm pt}$, top row) and different projectile-fireball friction scaling ($\xi_{\rm f}$, bottom row). The other friction settings are $\xi_{\rm f}=0.1$ for the top row ($\xi_{\rm pt}$ scan) and $\xi_{\rm pt}=1.0$ for the bottom row ($\xi_{\rm f}$ scan), and $\alpha=0.7$, $\beta=0.1$ for all plots. The data are by the PHOBOS~\cite{PHOBOS:2010eyu} and the NA49~\cite{NA49:1998gaz} collaborations.}\label{fig:xi-scan}
\end{figure*}

The charge transfer friction model aims to amend the IMS model by tying the assignment of scattered particles to fluids more closely to the idea of separation in momentum space. Accordingly, outgoing nucleons can also be assigned to the fireball. 
At first sight it looks like the charge transfer friction is a less suitable friction model, as it fails to reproduce the data. CT friction produces way less entropy than Csernai and IMS frictions, and consequently the multiplicity at midrapidity is way lower. Furthermore, baryon stopping is weaker than in the IMS model, and the peaks farther apart than in the data. However, as seen later this ``failure" leaves room for entropy production due to dissipation in individual fluids. Before discussing the role of dissipation, it is instructive to explore how different parameters of the charge transfer friction affect the final hadronic observables.

The charge transfer friction introduces two new parameters, $\alpha$ and $\beta$. The $\beta$ parameter sets a boundary in relative rapidity for the outgoing nucleons from the projectile-target scattering to go into either the original projectile/target fluids or to the fireball fluid. As shown in the top row of Fig.~\ref{fig:alpha-beta-scan}, with increasing $\beta$ the pseudorapidity distribution of all charged hadrons becomes somewhat narrower and develops a heightened plateau at mid-rapidity, because more baryons are being assigned to the fireball fluid. Correspondingly, the double peaks move farther apart and a third central peak appears in the rapidity distribution of net protons and becomes dominant at large $\beta$, which is not consistent with the experimental data at $\sqrts=19.6\,$GeV.

The $\alpha$ parameter controls how much energy is transferred to the fireball in nucleon-nucleon scattering, and how much of it is kept to heat up projectile and target. As shown in the bottom row of Fig.~\ref{fig:alpha-beta-scan}, with increasing $\alpha$, the charged hadron distribution becomes flatter in pseudorapidity, whereas  the mid-rapidity dip in the distribution of net protons becomes more pronounced. The larger the value of $\alpha$, the more energy is moved to the fireball due to projectile-target friction, which also increases the entropy of the fireball. Therefore, the larger the $\alpha$, the larger the multiplicity at midrapidity. The $\alpha$ parameter also regulates the loss of energy of the projectile and target fluids relative to the loss of their momenta. At a larger value of $\alpha$, more energy is lost compared to momentum. This leads to weaker deceleration of baryon-rich fluids, which results in a more pronounced double peak in the net-proton distribution. On the other hand, at very low $\alpha$, projectile and target are slowed down so much that unification is triggered for a significant portion of these baryon rich fluids and thus a large fraction of baryons is transferred to the fireball, leading to a third peak. 

Besides the new parameters $\alpha$ and $\beta$ of the charge transfer friction model, all models contain the parameters $\xi_{\rm pt}$ and $\xi_{\rm f}$ controlling the overall strength of the projectile-target and fireball frictions. The top row of Fig.~\ref{fig:xi-scan} shows the results of a scan of the $\xi_{pt}$-parameter, which has strong effects on both distributions. For low values of $\xi_{\rm pt}$, little energy and momentum is transferred to the fireball, resulting in a small charged hadron multiplicity far undershooting the experimental data and slightly broader pseudorapidity distribution than experimentally observed. As projectile and target experience only little deceleration, the two peaks in the net-proton distribution are far apart in rapidity, close to the ingoing beam rapidities. In contrast, with increasing $\xi_{\rm pt}$, a hotter fireball results in significantly higher charged hadron production, while the two net-proton peaks become wider and move closer in rapidity eventually merging to a single peak. Choosing $\xi_{\rm pt}=2.0$ yields the best data agreement for the charged hadron distribution, but completely eliminates the double peak structure in the net-proton distribution.
 
\begin{figure*}[ht]
 \centering
 \includegraphics[width=0.49\textwidth]{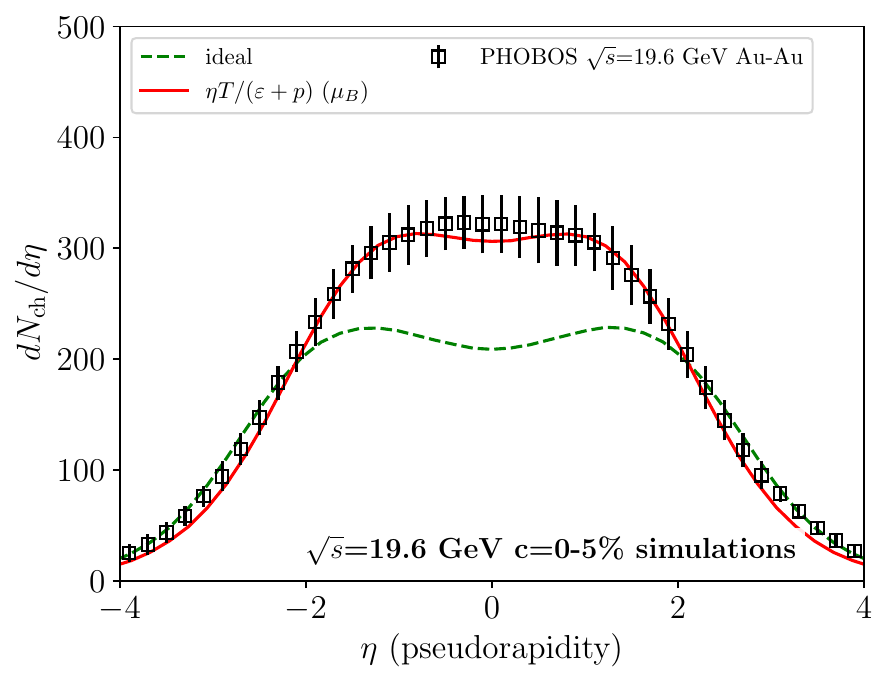}
 \includegraphics[width=0.49\textwidth]{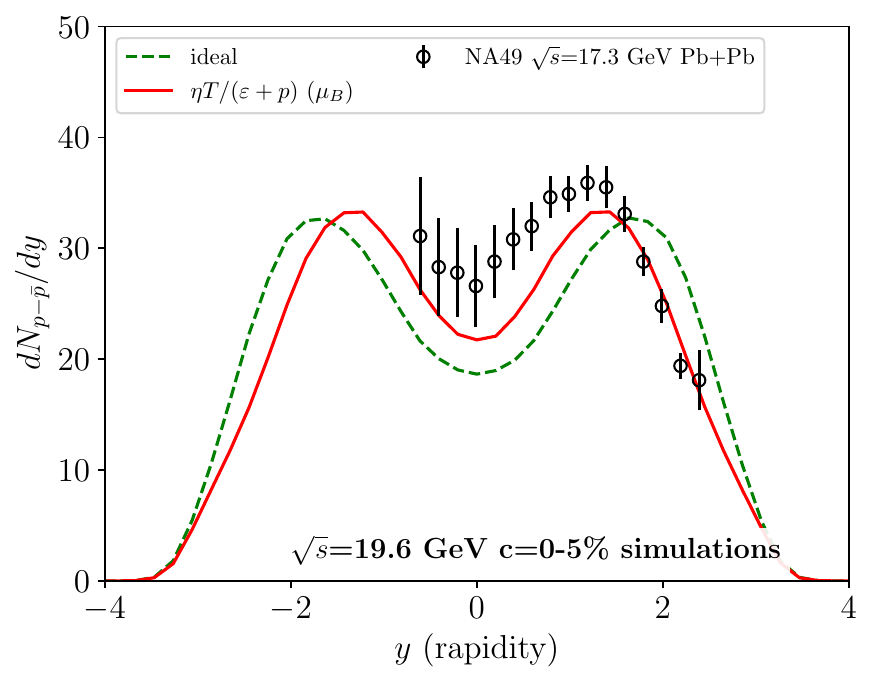}
 \caption{Comparison of results from multi-fluid simulations with ideal fluids to fluids with baryon chemical potential dependent shear viscosity from Eq.~\eqref{eq:etaS}. Friction settings $\alpha=0.7$, $\beta=0.1$, $\xi_{\rm pt} = 1.0$, $\xi_{\rm f}=0.1$ are used for this comparison.The data are by the PHOBOS~\cite{PHOBOS:2010eyu} and the NA49~\cite{NA49:1998gaz} collaborations. }\label{fig:id-visc}
\end{figure*}

The results of a scan in $\xi_{\rm f}$ are displayed in the bottom row of Fig.~\ref{fig:xi-scan}. This parameter scales the fireball friction, which in this model moves additional conserved charges from projectile and target to the fireball once the latter has already accumulated a significant density. This does not seem to have a huge effect on total charged hadron multiplicity, but it does modulate their distribution. With increased $\xi_{\rm f}$, a stronger double peak structure appears, suggesting that the outskirts of the fireball fluid are accelerated when absorbing the contents of the projectile and target due to fireball friction. This could also explain the shoulders that appear in the net-proton distribution. Here, the main double peak structure is washed out with increasing $\xi_{\rm f}$ and pushed to midrapidity as projectile and target are drained more.

Collecting observations from all parameter scans, one can spot a general tension between the reproduction of the experimental charged hadron pseudorapidity distribution $dN_{\rm ch}/d\eta$ and net-proton rapidity distribution $dN_{p-\bar{p}}/dy$. In our test simulations we have observed that typically a variation of an input parameter that improves agreement with the midrapidity $dN_{\rm ch}/d\eta$ from the data also makes the distribution of net protons narrower or eliminates the double peak structure that was measured by NA49 altogether. Therefore, in simulations of central gold-gold collisions at $\sqrts=19.6\,$GeV, parameter combinations that resulted in a reasonable reproduction of experimental $dN_{\rm ch}/d\eta$ led to a shape of $dN_{p-\bar{p}}/dy$ that would be incompatible with the data. As shown in Fig.~\ref{fig:FMs}, even our optimal parameter combination $\alpha = 0.7$, $\beta = 0.1$, $\xi_{\rm pt} = 1.0$, $\xi_{\rm f} = 0.1$, fails to reproduce the charged hadron multiplicity at midrapidity. However, it turned out that this ``failure" is not a bug, but a feature.

\section{Shear viscosity}

The multi-fluid model is constructed to describe the highly non-equilibrated system at the early stages of evolution, but the construction using multiple fluids is insufficient to describe dissipation within the system. As mentioned, so far the multi-fluid calculations have assumed ideal fluid behaviour of each constituent fluid. The described underestimation of charged hadron multiplicity in model scenarios that provide reasonable predictions for the net-proton distribution could well be a consequence of neglecting dissipation, i.e.\ of the ideal hydrodynamic description of each constituent fluid. 

At $\sqrts=19.6\,$GeV, the created fireball fluid is dense and exhibits a strong  longitudinal expansion. Here, shear viscosity acts in the same way as in one-fluid description of collisions at much larger collision energies: it reduces the anisotropy of the expansion by inhibiting the longitudinal expansion and enhancing the transverse one. As a result, more energy of the fireball fluid is trapped in mid-spacetime rapidity slices, leading to stronger expansion in the transverse direction, larger final-state volume of the system and overall more entropy production, leading to higher produced hadron multiplicity.

The results of our simulations including shear viscosity are shown in Fig.~\ref{fig:id-visc}. As expected, the charged hadron multiplicity is significantly increased and reproduces the data nicely. At the same time, the shear viscosity does not seem to affect the distribution of net protons too much. The main effect is a slight narrowing of the $dN_{p-\bar{p}}/dy$ distribution, which agrees with the idea of weaker longitudinal expansion. Overall, the inclusion of shear viscosity in the description reduces the tension between the reproduction of observed charged hadron pseudorapidity distribution $dN_{\rm ch}/d\eta$ and net-proton rapidity distribution $dN_{p-\bar{p}}/dy$.

\begin{figure*}[ht]
 \centering
 \includegraphics[width=0.49\textwidth]{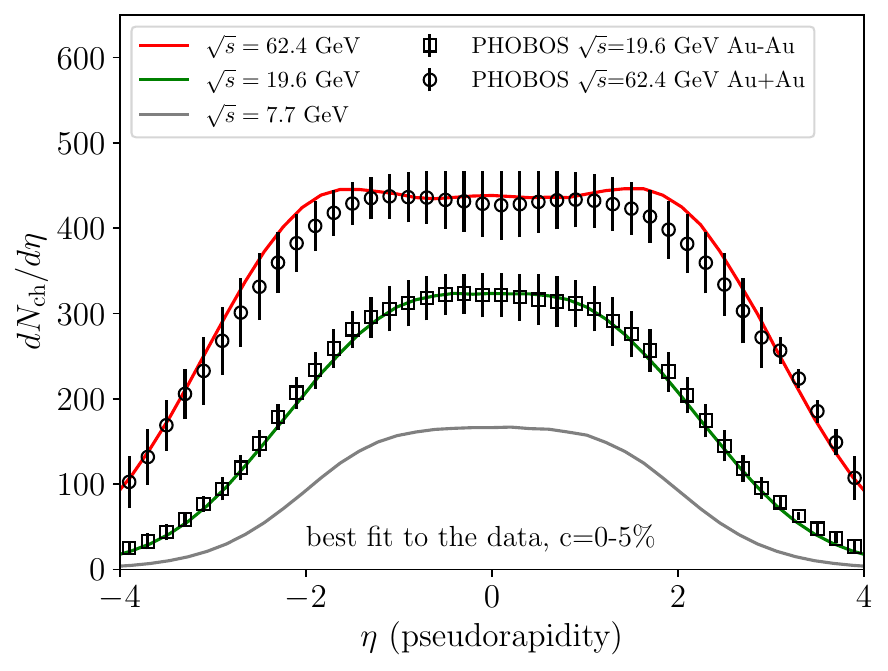}
 \includegraphics[width=0.49\textwidth]{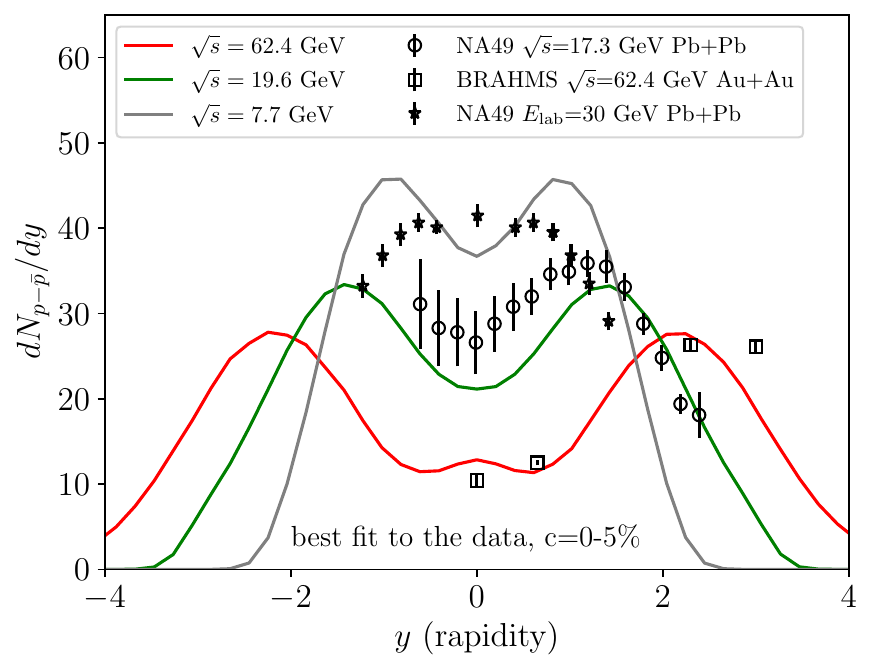}
 \caption{Best fit to the experimental data at three different collision energies, achieved with the friction terms settings $\alpha=0.8$, $\beta=0.1$, $\xi_{f\alpha}=0.1$, $\xi_{pt}=1.0$. For $\sqrts=62.4$~GeV, an alternation $\alpha=0.75$, $\xi_{pt}=0.8$ is used. Fluids are evolved with temperature- and baryon chemical potential dependent shear viscosity from Eq.~\eqref{eq:etaS}.  Data are by the PHOBOS~\cite{PHOBOS:2010eyu}, NA49~\cite{NA49:1998gaz,Blume:2007kw} and BRAHMS~\cite{BRAHMS:2009wlg} collaborations.}\label{fig:best-fit}
\end{figure*}

\section{Further data comparison}

Up to this point, to gain insight in the roles of each of the model parameters in terms of physics and phenomenology, we concentrated on the Au+Au collisions at $\sqrt{s_{NN}}=19.6\,$GeV. Next, to test the predictive power of our model, we calculated the charged hadron pseudorapidity distributions $dN_{\rm ch}/d\eta$, net-proton rapidity distributions $dN_{p-\bar{p}}/dy$, identified particle $p_T$-distributions ($\pi^-$, $K^-$ and protons) and $p_T$-differential elliptic flow $v_2(p_T)$ of charged hadrons for the best parameter combinations at four different collision energies: $7.7$, $19.6$, $39$ and $62.4\,$GeV.

The pseudorapidity and rapidity distributions are shown in Fig.~\ref{fig:best-fit}. Data for the net-proton rapidity distribution is sparse, but for $7.7$ and $62.4\,$GeV it has been measured by the NA49~\cite{NA49:1998gaz,Blume:2007kw} and BRAHMS~\cite{BRAHMS:2009wlg} collaborations. Additionally, data for the charged hadron distribution in pseudorapidity is available at $62.4\,$GeV owing to PHOBOS~\cite{PHOBOS:2010eyu}. The plots show results for the parameter combination that best described experimental data. For $7.7\,$GeV and $19.6\,$GeV, these parameters are the same, and mentioned above, but for $62.4\,$GeV $\alpha = 0.75$ and $\xi_{\rm pt} = 0.8$ led to better fit. We note that in previous works~\cite{Ivanov:2013wha,Cimerman:2023hjw} using IMS friction, the parameters $\xi_{\rm pt}$ and $\xi_{\rm f}$ were modeled to depend on the CMS energy of the microscopic scattering processes, meaning they would change throughout the evolution. We use here a much weaker dependence only on the $\sqrts$ of the entire nucleus-nucleus collision system. Our  longitudinal distributions are in reasonable agreement with experimental data at all three collision energies. The model correctly reproduces the increase and widening of the charged hadron distribution with increasing energy. It also adequately reproduces in the net-proton distribution a transition from having almost no double peak structure at $7.7\,$GeV to exhibiting two strongly separated peaks at $62.4\,$GeV.

\begin{figure*}[ht]
 \centering
 \includegraphics[width=0.49\textwidth]{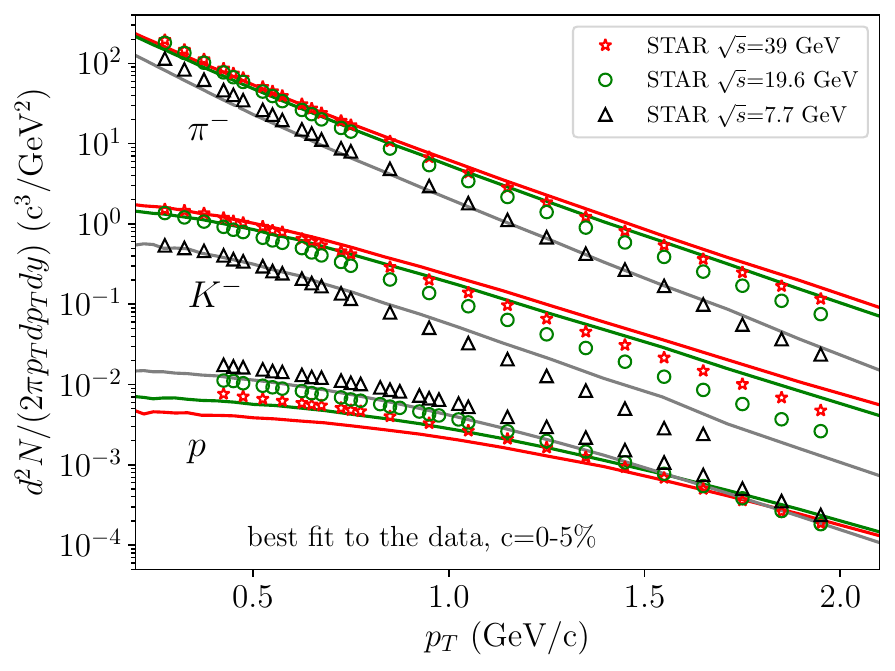}
 \includegraphics[width=0.49\textwidth]{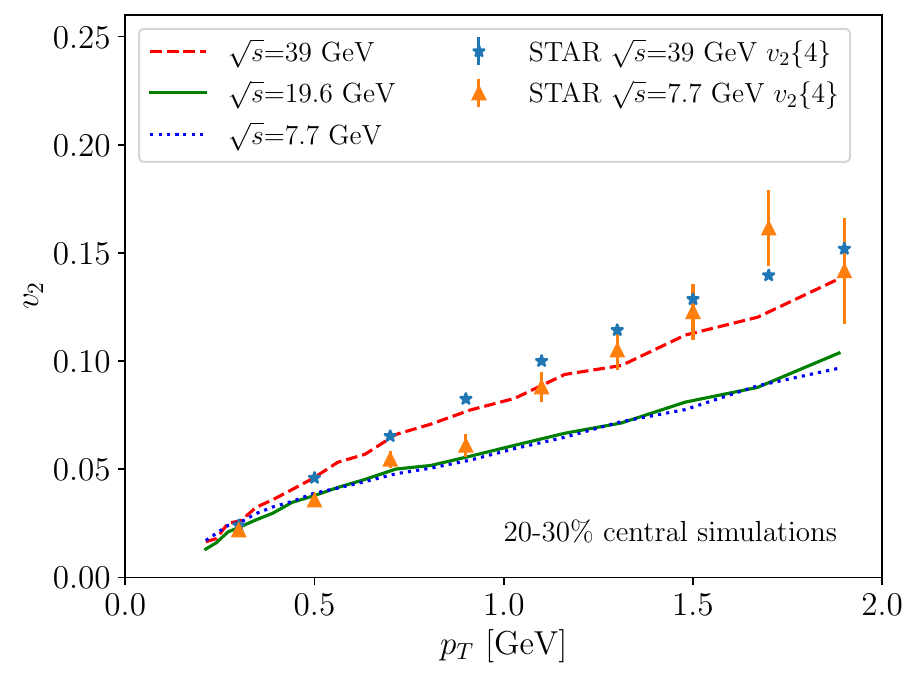}
 \caption{(Left) Transverse momentum distributions of negative pions, negative kaons and protons, computed from simulations of 0-5\% central Au-Au collisions and (right) elliptic flow of all charged hadrons as a function of transverse momentum, computed from simulations of 20-30\% central Au-Au collisions, both with the same settings as described in Fig.~\ref{fig:best-fit}. The experimental data by the STAR collaboration is taken from \cite{STAR:2017sal,STAR:2012och}.}\label{fig:best-fit-pt-v2}
\end{figure*}

In Fig.~\ref{fig:best-fit-pt-v2} the $p_T$-distributions of identified hadrons and the $p_T$-differential elliptic flow $v_2(p_T)$ of charged hadrons at $\sqrt{s_{\rm NN}} = 7.7$, 19.6 and 39 GeV collision energies are shown and compared with the data by the STAR collaboration~\cite{STAR:2017sal,STAR:2012och}. 

First, our $p_T$-distributions in the left plot of Fig.~\ref{fig:best-fit-pt-v2} mostly show agreement at small $p_T$ but tend to be too hard which may indicate the need for bulk viscosity and/or more sophisticated temperature/chemical potential dependence of the shear viscosity coefficient $\eta$. Pion $p_T$ spectra from STAR have weak decay contributions subtracted, whereas proton ones do not. In our simulations, weak decays are switched off, which makes a more consistent comparison for pions but less consistent one for protons. The weak decay contributions explain the difference in the proton yields between our calculations and the experiment.

Second, in~\cite{Cimerman:2023hjw} the calculated elliptic flow clearly overshot expectations, but inclusion of viscosity has brought it down to the vicinity of the data, as seen in the right plot of Fig.~\ref{fig:best-fit-pt-v2}. We reproduce the data up to $p_T \approx 1$ GeV, but at larger values of $p_T$ our calculation suppresses $v_2$ too much, which again may indicate the need of bulk viscosity or improved shear viscosity parametrization $\eta(T,\mu)$. Nevertheless, it must be remembered that all the parameters of our model were fixed by comparison to rapidity and pseudorapidity distributions only, and our $p_T$ and elliptic flow results are predictions. From that point of view, our results are reasonable, and provide a good basis for further refinement of our model.

\section{Conclusion}

In this work, we have introduced a new model for the friction between projectile and target in multi-fluid dynamics, which we dubbed charge transfer friction. This model interpolates between two existing friction models: Csernai friction moving all scattered particles to the fireball and IMS friction keeping all baryon charge in projectile and target. Csernai friction transports baryon charge in the same fashion as energy and therefore cannot describe baryon transparency. IMS friction improves on this since baryon charge moves with the fast target and projectile fluids, but might decelerate them too much. Charge transfer friction introduces further parameters that allows more flexibility in reproducing the net-proton distribution, but seems to underestimate charged hadron production at midrapidity.

We have performed scans of the four parameters of charge transfer friction to develop an understanding of their role in the physics at play. All parameters affect both particle production and baryon stopping via the amount of energy in the fireball, deceleration of target and projectile, expansion of the fireball, and/or the hadro-chemistry of the fluids. We did not find a parameter combination that would result in a simultaneous agreement with the experimental data for both the pseudorapidity density of all charged hadrons and the rapidity density of net protons. Cases that correctly reproduce the net-proton distribution would persistently underestimate charged hadron multiplicity. This leaves room for shear viscosity in the internal evolution of the fluids to increase entropy production. As we observed, inclusion of shear led to good agreement with data even across collision energies.

Together with the previously observed overestimation of elliptic flow~\cite{Cimerman:2023hjw}, it now seems clear that dissipative effects are important in multi-fluid dynamical modeling of low energy collisions, which opens up several further research questions for future work. In addition to shear, the model should include bulk viscosity and charge diffusion as further dissipative effects. Furthermore, with the inclusion of dissipative quantities, one should also describe how they are affected by inter-fluid friction in a way that is consistent with the transfer of conserved quantities.

\acknowledgements

CW has received funding from the European Research Council (ERC) under the European Union’s Horizon 2020 research and innovation programme (grant number: 101089093 / project acronym: High-TheQ). Views and opinions expressed are however those of the authors only and do not necessarily reflect those of the European Union or the European Research Council. Neither the European Union nor the granting authority can be held responsible for them.
IK acknowledges support by the Czech Science Foundation under project No.~25-16877S.
PH was supported by the program Excellence Initiative–Research University of the University of Wroc\l{}aw of the Ministry of Education and Science. Computational resources were provided by the e-INFRA CZ project (ID:90254), supported by the Ministry of Education, Youth and Sports of the Czech Republic, and by Wroclaw Centre for Networking and Supercomputing (\url{http:// wcss.pl}).

\section*{Data availability}

The data that support the findings of this article are
openly available~\cite{Zenodo}.

\appendix

\section{Derivation of friction terms}\label{app:deriving_friction}

The friction terms describing the transfer of energy and momentum between fluids can be derived from an underlying microscopic theory. We follow Refs.~\cite{Satarov:1990uc,Mishustin:1991sp,Ivanov:2005yw} in deriving them from kinetic theory, but we will make different modeling assumptions and point out where they play a role.

In kinetic theory, the system can be divided into different components on the level of phase space distributions, $f_I(t,\mathbf{x},\mathbf{p})$. Their evolution is then given by a set of Boltzmann equations that are coupled via the collision kernels $C_{JK\to I X}[f_J,f_K]$ describing the particles ending up in component $I$ due to scatterings of particles from the components $J$ and $K$ and $C_{IJ\to X}[f_I,f_J]$ describing the particles of component $I$  scattering with particles from component $J$:
\begin{align}
    p^\mu\partial_\mu f_I&=C_I[\{f_J\}]=\sum_J C_{JJ\to I X}[f_J]\label{eq:BoltzmannI}\\
    &\quad+ \frac{1}{2}\sum_{J\neq K}C_{JK\to I X}[f_J,f_K]-\sum_J C_{IJ\to X}[f_I,f_J].\nonumber
\end{align}

The energy-momentum tensor $T_I^{\mu\nu}$ and baryon current $N_I^\mu$ of the component $I$ can be derived from the corresponding phase space distribution as
\begin{align}
    T_I^{\mu\nu}&=\int\d\Gamma_I~{p_I^\mu p_I^\nu}f_I\;,\\
    N_I^{\mu}&=\int\d\Gamma_I~ {p_I^\mu } f_I\;,
\end{align}
where we defined $\d\Gamma_I={\d^3p_I}/{[(2\pi)^3p_I^0]}$. Thus, taking appropriate moments of the Boltzmann equation~\eqref{eq:BoltzmannI} allows to derive evolution equations for $T_I^{\mu\nu}$ and $N_I^{\mu}$:
\begin{align}
    \partial_\mu T_I^{\mu\nu}&=\int\d\Gamma_I ~{ p_I^\nu} C_I\;,\label{eq:Tmunuevo}\\
    \partial_\mu N_I^{\mu}&=\int\d\Gamma_I~ C_I.\label{eq:Nmuevo}
\end{align}

Notice that one is free to choose what particles are considered to be part of what component. The multi-fluid Ansatz separates particles by their momentum into projectile and target at high rapidities and the fireball at midrapidity, describing the internal evolution of each individual component via fluid dynamics. This means that the effect of interactions within fluid $I$ ($C_{II\to IX}$ and $C_{II\to X}$) are taken care of by its fluid dynamical description, which nets zero change in $T^{\mu\nu}_I$. This fluid dynamical picture also dictates that no particles from the same fluid $J$ can produce particles in another fluid $I$, $C_{JJ\to IX}=0$ for $J\neq I$. All other collision kernels will via Eq.s~\eqref{eq:Tmunuevo} and~\eqref{eq:Nmuevo} generate friction contributions as
\begin{align}
    \partial_\mu T_I^{\mu\nu}=F_I^\nu=&\frac{1}{2}\sum_{J\neq K} \int\d\Gamma_I~ { p_I^\nu} C_{JK\to IX}[f_J,f_K]\\
    &-\sum_{J\neq I}\int\d\Gamma_I~ p_I^\nu C_{IJ\to X}[f_I,f_J]\;,\nonumber\\
    \partial_\mu N_I^{\mu}=R_I=&\frac{1}{2}\sum_{J\neq K} \int\d\Gamma_I~ C_{JK\to IX}[f_J,f_K]\\
     &-\sum_{J\neq I}\int\d\Gamma_I~ C_{IJ\to X}[f_I,f_J]\;.\nonumber
\end{align}

The collision kernels can now be expressed in terms of scattering cross-sections. Each component is considered to be made up of one type of particle, resulting in one specific cross-section. In the case that the fluids are mixtures, each fluid has to be considered to be made up of multiple components, but we will not model that here. We will also consider classical statistics, meaning that no quantum statistical factors occur in the collision kernel:
\begin{align}
    C_{JK\to IX}[f_J,f_K]=&\int\d\Gamma_J~f_J(p_J)\int\d\Gamma_K ~f_K(p_K)\\
    &\sqrt{\left(p_{J,\mu}p_K^\mu\right)^2-m_J^2m_K^2} ~\frac{\d\sigma_{JK\to IX}}{\d\Gamma_I},\nonumber\\
     C_{IJ\to X}[f_I,f_J]=&f_I(p_I)\int\d\Gamma_J~f_J(p_J)\\
     &\sqrt{\left(p_{I,\mu}p_J^\mu\right)^2-m_I^2m_J^2} ~\sigma_{IJ\to X}.\nonumber
\end{align}

Up to this point, the equations are general. But in order to arrive at explicit formulae, the cross-sections $\sigma_{JK\to IX}$ have to be modeled, which corresponds to choosing to which of the fluids outgoing particles from the microscopic scatterings are supposed to move.

For the model used in~\cite{Ivanov:2005yw,Ivanov:2013wha} (and~\cite{Satarov:1990uc,Mishustin:1991sp} without fireball friction), this set of modeling choices can be summarised as follows. The fast projectile and target fluids are composed of nucleons, while the fireball at midrapidity is made of pions. Therefore, all outgoing nucleons from nucleon-nucleon scattering will go to either projectile or target depending on whether it was emitted in the forward or backward direction in the CMS frame of the collision, while all other particles produced in inelastic collisions go to the fireball. When a particle from projectile or target scatters with a particle from the fireball, the result is assumed to be resonance formation, with an outgoing momentum close to that of the ingoing nucleon, and the outgoing  resonance and its daughter particles will stay with the fluid that the nucleon came from. This corresponds to $\d\sigma_{pt\to pX}/\d\Gamma_p=\theta(\tilde{p}_p)~\d\sigma_{NN\to NX}/\d\Gamma_p$, $\sigma_{pt\to X}=\sigma_{NN\to X}$, $\sigma_{pf\to X}=\sigma_{\pi N\to X}$, $\d\sigma_{pf\to p X}/\d\Gamma_p=\d\sigma_{\pi N \to R}/\d\Gamma_p$ and $\d\sigma_{pf \to fX}/\d\Gamma_f=\d\sigma_{pf \to tX}/\d\Gamma_t=0$.
The remaining friction terms are fixed by symmetry $p\leftrightarrow t$ as well as energy-momentum and baryon number conservation. Under the assumption that target and projectile have no thermal spread, $f_{p/t}\propto \delta^{(3)}(p_{p/t}-m_N u_{p/t})$, and due to forward-backward symmetry of $\d\sigma_{NN\to NX}/\d\Gamma$, the momentum integrals can be evaluated straightforwardly without differential input for the cross-sections, leading to the IMS friction terms stated in Sec.~\ref{sec:model}.

Motivated by the goal of studying the QCD equation of state at nonzero $\mu_B$, we want to employ a slightly more realistic friction model that allows baryons to enter the fireball. In particular, we will make three distinctions for the outgoing nucleons from scatterings between target and projectile. In the center of mass frame, for incoming nucleons at rapidities $0<{y}_p=-{y}_t$, the outgoing nucleon is modeled to move to the fireball if its rapidity is $|y|<\beta y_p$ and otherwise to target or projectile depending on the sign of $y$, where $\beta\in[0,1]$ is a free model parameter. With this choice, we will now need differential information on $\d\sigma_{NN\to NX}/\d\Gamma$.

In order to fix on top of baryon transfer also the energy balance of the target-projectile friction, technically we would need full information on $\d\sigma_{NN\to NX}/(\d y\d E)$, which is not available. Instead, we make the simplifying assumption that a fixed
fraction of the part of the incoming energy $E_0$ that is not transferred to
longitudinal momentum of the outgoing nucleon, $m_N \cosh(y)$, will be
transferred to the fireball, while the rest remains
with the outgoing nucleon as transverse momentum. In other words,
\begin{align}
    E_{\rm f}&=\alpha[E_0-m_N\cosh(y)]\;,\\
    E&=E_0-E_{\rm f}\;,
\end{align}
where $E$ is the outgoing nucleon's energy, $E_{\rm f}$ is the energy that goes to the fireball, and $\alpha\in[0,1]$ is a second free model parameter.

Furthermore, we assume for $\d\sigma_{NN\to NX}/\d y$ a functional form that can approximately describe the data but also allows to solve the occurring momentum integrals analytically:
\begin{align}
    \frac{\d\sigma_{NN\to NX}}{\d y}\propto1+6\frac{\cosh(10y)}{\cosh(10y_p)}.
\end{align}

This form is inspired by the empirical observation that in terms of the squared four-momentum transfer $t$, the elastic nucleon-nucleon cross-section decays exponentially from both ends of its spectrum, $\d\sigma_{NN\to NN}/\d t\propto e^{-bt}+e^{-b(t_{\rm max}-t)}$~\cite{Cugnon:1996kh}, which can be written as a hyperbolic cosine, and is also adjusted to reproduce the plot of $\d\sigma_{NN \to NX}/\d y$ in Fig. 1 of~\cite{Satarov:1990uc}.

When evaluating the momentum integration in the friction terms with this form of the differential cross-section, the following integrals will occur.
\begin{align}
     A=&\int_0^{y_p} 1+6\frac{\cosh(10y)}{\cosh(10y_p)} \d y= y_p+\frac{3}{5}\\
     B=&\int_0^{\beta y_p} 1+6\frac{\cosh(10y)}{\cosh(10y_p)} \d y= \beta y_p+\frac{3}{5}\frac{\sinh(10\beta y_p}{\cosh(10 y_p)}\\
     C=&\int_{\beta y_p}^{y_p} \left(1+6\frac{\cosh(10y)}{\cosh(10y_p)}\right) \frac{\tanh(y)}{\tanh(y_p)} \d y\\
     =&\frac{\log\left(\frac{\cosh(y_p)}{\cosh(\beta y_p)}\right)}{\tanh(y_p)}\left(1-\frac{6}{\cosh(10y_p)}\right)\nonumber\\
     &+\frac{1}{\cosh(10y_p)\tanh(y_p)} \Big[6\cosh(2y_p)\nonumber\\
     &+6\cosh(2\beta y_p)-3\cosh(4y_p)+3\cosh(4\beta y_p)\nonumber\\
     &+2\cosh(6y_p)-2\cosh(6\beta y_p)-\frac{3}{2}\cosh(8y_p)\nonumber\\
     &+\frac{3}{2}\cosh(8\beta y_p)+\frac{3}{5}\cosh(10y_p)-\frac{3}{5}\cosh(10\beta y_p)\Big]\nonumber
\end{align}
\begin{align}
          D=&\int_{\beta y_p}^{y_p} \left(1+6\frac{\cosh(10y)}{\cosh(10y_p)}\right) \frac{\cosh(y)}{\cosh(y_p)} \d y\\
     =&\frac{\sinh(y_p)-\sinh(\beta y_p)}{\cosh(y_p)}+\frac{1}{3}\frac{\sinh(9y_p)-\sinh(9\beta y_p)}{\cosh(y_p)\cosh(10 y_p)}\nonumber\\
     &+\frac{3}{11}\frac{\sinh(11 y_p)-\sinh(11\beta y_p)}{\cosh(y_p)\cosh(10 y_p)}\nonumber\\
     F=&\int_{\beta y_p}^{y_p} \left(1+6\frac{\cosh(10y)}{\cosh(10y_p)}\right) \frac{\sinh(y)}{\sinh(y_p)} \d y\\
     =&\frac{\cosh(y_p)-\cosh(\beta y_p)}{\sinh(y_p)}-\frac{1}{3}\frac{\cosh(9y_p)-\cosh(9\beta y_p)}{\sinh(y_p)\cosh(10 y_p)}\nonumber\\
     &+\frac{3}{11}\frac{\cosh(11 y_p)-\cosh(11\beta y_p)}{\sinh(y_p)\cosh(10 y_p)}\nonumber
\end{align}

The normalization of $\d\sigma_{NN\to NX}/\d y$ is of course fixed by the total cross-section:
\begin{align}
    \frac{\d \sigma_{NN\to NX}}{dy}&=\frac{\sigma_{NN\to NX}}{2A} \left[ 1+6\frac{\cosh(10y)}{\cosh(10y_p)}\right].
\end{align}

Now we can start computing the friction terms. The change in baryon number in the projectile due to projectile-target interaction is given by
\begin{align}
R_{B,pt}=&\int\d\Gamma_p ~C_{pt\to X}[f_p,f_t]\\
=&\int \d \Gamma_p f_p \int \d \Gamma_t f_t \sqrt{(p_p^\mu p_{t,\mu})^2-m_N^4}\\
&\times\Bigg[-\underbrace{\int_{\bar{v}_p}\d\Gamma \frac{\d\sigma_{NN\rightarrow NX}}{\d\Gamma}}_{=\bar{\sigma}_R(s)} \Bigg]\;.\nonumber
\end{align}
Here we have defined $\bar{v}_p=\{0<y<\beta y_p\}$ and
\begin{align}
    \bar{\sigma}_R(s)&=\frac{\sigma_{NN\to NX}(s)}{2}\frac{B}{A}\;.
\end{align}

Similarly, the change of energy and momentum due to projectile-target interaction evaluates to
\begin{align}
    &{F}_{pt}^\nu=\int \d \Gamma ~p^\nu~ (C_{pt\to p'X}[f_p,f_t]-C_{pt\to X}[f_p,f_t])\\
    &=\int \d \Gamma_p f_p \int \d \Gamma_t f_t \sqrt{(p_p^\mu p_{t,\mu})^2-m_N^4}\\
    &\Bigg[\underbrace{\int_{\bar{w}_p}\d\Gamma ~(p^\nu-p_p^\nu)\frac{\sigma_{NN\rightarrow NX}}{\d\Gamma}}_{=(*)^\nu} -\underbrace{\int_{\bar{v}_p}\d \Gamma \frac{\d\sigma_{NN\rightarrow NX}}{\d\Gamma} p_p^\nu}_{=\bar{\sigma}_R(s)p_p^\nu}\Bigg]\;,\nonumber
\end{align}
where $\bar{w}_p=\{\beta y_p<y<y_p\}$. We can make further progress by rewriting $p$ as a linear combination of $p_p+p_t$ and $p_p-p_t$, which yields
\begin{align}
    (*)^\nu=-\frac{1}{2}[\bar{\sigma}_P(s)(p_p^\nu-p_{t}^\nu)+\bar{\sigma}_E(s)(p_p^\nu+p_{t}^\nu)]\;,
\end{align}
where
\begin{align}
    \bar{\sigma}_E(s)&=\int_{\bar{w}_p}\d\Gamma \frac{\d\sigma_{NN\rightarrow NX}}{\d\Gamma} \left( 1-\frac{E}{E_0} \right)\\
    &=\int_{\beta y_p}^{y_p}dy  \frac{\d\sigma_{NN\rightarrow NX}}{dy} \left[1-\frac{\cosh(y)}{\cosh(y_p)}\right]\\
    &=\frac{\sigma_{NN\rightarrow NX}(s)}{2}\frac{A-B-D}{A}\;,\\
    \bar{\sigma}_P(s)&=\int_{\bar{w}_p}\d\Gamma \frac{\d\sigma_{NN\rightarrow NX}}{\d\Gamma} \left( 1-\frac{p_L}{p_0} \right)\\
    &=\int_{\beta y_p}^{y_p}dy  \frac{\d\sigma_{NN\rightarrow NX}}{dy} \left[1-\frac{E}{E_0}\frac{\tanh(y)}{\tanh(y_p)}\right]\\
    &=\int_{\beta y_p}^{y_p}dy  \frac{\d\sigma_{NN\rightarrow NX}}{dy}\\
    &\quad\times\left\{1-\left[1-\alpha+\alpha\frac{\cosh(y)}{\cosh(y_p)}\right]\frac{\tanh(y)}{\tanh(y_p)}\right\}\nonumber\\
    &=\frac{\sigma_{NN\rightarrow NX}(s)}{2}\left(\frac{A-B}{A}-(1-\alpha)\frac{C}{A}-\alpha\frac{F}{A}\right)\;.
\end{align}
Finally, with the assumption of no thermal spread, $f_{p/t}\propto \delta^{(3)}(p_{p/t}-m_N u_{p/t})$, the remaining momentum integrals become trivial and we obtain formulae for the friction terms that depend on $u^\mu_{p/t}$ and $n_{p/t}=m_N\int\d\Gamma f_{p/t}$. Defining $s_{pt}=m_N^2(u_p^\mu+u_t^\mu)^2$ and $V^{pt}_{\rm rel}=\sqrt{(u_{p,\mu}u_t^\mu)^2-1}$, we can write them as
\begin{align}
     R_{B,pt}=&-n_p n_t V^{pt}_{\rm rel}u^\nu_p\bar{\sigma}_R(s_{pt}),\\
    F_{pt}^\nu=&-n_p n_t m_N V^{pt}_{\rm rel}\Big[\frac{1}{2}(u_p^\nu-u_{t}^\nu)\bar{\sigma}_P(s_{pt})\\
    &+\frac{1}{2}(u_p^\nu+u_{t}^\nu)\bar{\sigma}_E(s_{pt})+u^\nu_p\bar{\sigma}_R(s_{pt})\Big]\;.\nonumber
\end{align}
Of course, the same equations with $p\leftrightarrow t$ describe the changes of the conserved quantities in the target fluid.

\section{Fireball friction with thermal spread}\label{app:fireball_friction}

The friction of projectile or target with the fireball fluid was first considered in~\cite{Ivanov:2005yw}. The functional form of the friction term derived there was later also used in~\cite{
Ivanov:2013wha,Ivanov:2013yqa,Ivanov:2013yla,Batyuk:2016qmb,Kozhevnikova:2020bdb,Cimerman:2023hjw}. Since in this model all particles coming out of interactions between projectile and fireball are considered to move to the projectile, the friction is fully determined by the loss term of the fireball and energy-momentum conservation. Again, there is no transfer of baryon charge. Thermal spread of both fluids was neglected in the derivation in order to arrive at analytical results for the momentum integrals. However, instead of expressing it via density, a factor $1\approx p^0_f/(m_\pi u_f^0)$ was added to the integral over momenta of fireball particles in order to express it via the corresponding energy-momentum tensor.
\begin{align}
    &F_{pf}^\nu=\int\d\Gamma_f~ p_f^\nu~(-C_{pf\to X}[f_p,f_f]) \\
    &=\int\d\Gamma_pf_p\int\d\Gamma_f~p_f^\nu f_f \sqrt{(p_{\mu,f}p^\mu_p)-m_N^2m_\pi^2}~\sigma_{N\pi\to X}(s)\\
    &\approx V_{\rm rel}^{pf}~\sigma_{N\pi\to X}(s_{pf})~ n_p \frac{T_f^{0\nu}}{u_f^0}
\end{align}

Neglecting thermal spread is a drastic approximation, especially for the fireball consisting mostly of hot pions. Thus, we want to derive new formulae that go beyond this approximation by including at least the thermal spread of the fireball. However, some of the occurring integrals will not be analytically solvable and have to be tabulated as a function of fireball temperature and relative velocity. In principle, generalization to including thermal spread also of the projectile or even the baryon content of the fireball is straightforward, but leads to more parametric dependencies in which the friction would have to be tabulated.

We will also keep the assumption that the fireball is made up exclusively of pions. To simplify the calculation of the friction, we perform it in the fireball fluid's restframe as denoted by a tilde for all corresponding quantities. In this frame, its phase space distribution depends only on the particles' energy. Specifically, for the number of pion degrees of freedom, $g_\pi=3$:
\begin{align}
    \tilde{f}_f=\frac{g_\pi}{\exp(\tilde{p}^0/T_f)-1}\;.
\end{align}

In contrast to the previous model, we will assign all outgoing particles from projectile-fireball interactions to the fireball. Then, the friction term can be computed straightforwardly from the loss term in the projectile:
\begin{align}
    &\tilde{F}_{pf}^\nu=\int\d\tilde{\Gamma}_p~\tilde{p}_p^\nu(-\tilde{C}_{pf\to X}[\tilde{f}_p,\tilde{f}_f])\\
    &=\int\d\tilde{\Gamma}_p~\tilde{p}_p^\nu \tilde{f}_p\int\d\tilde{\Gamma}_f\tilde{f}_f \sqrt{(\tilde{p}_{p,\mu}\tilde{p}_f^\mu)^2-m_\pi^2 m_N^2}~\sigma_{N\pi\to X}(s)\label{eq:fpfric_Lorentz}
\end{align}
and analogously for the baryon charge transfer $R_{B,p}$ without the factor $\tilde{p}_p^\nu$. After assuming no thermal spread in the projectile, the CMS squared energy $s$ can be expressed as a function of $\tilde{p}_0^f$ and the flow velocities.
\begin{align}
    s=({p}_f^\mu+{p}_p^\mu)^2=&m_N^2+m_\pi^2+2m_N u_{p,\mu}p_f^\mu\\
    \approx&m_N^2+m_\pi^2+2m_N \gamma_{pf} \tilde{p}_f^0\;,
\end{align}
where $\gamma_{pf}$ is the Lorentz $\gamma$-factor associated to the relative velocity of the projectile and fireball. The last step technically partly neglects thermal spread by dropping a term $-2m_N\gamma_{pf}v_{pf} p_{pf}\cos(\theta_{pf})$. The error in the friction from this approximation is larger for larger $T_f$ and $v_{pf}$. For $T_f\le 0.5\,\mathrm{GeV}/1\,\mathrm{GeV}$ and $v_{pf}\le 0.5/0.9$, it is below $3\%/10\%$.

The M\o ller flux factor can be rewritten as
\begin{align}
    \sqrt{(p_{p,\mu}p_f^\mu)^2-m_N^2m_\pi^2}=\frac{1}{2}\sqrt{(s-m_N^2-m_\pi^2)^2-4m_N^2m_\pi^2}
\end{align}
and thus has the same parametric dependencies as $s$. Now we can perform the integral over $p_p$ and the angles in $p_f$:
\begin{align}
    \tilde{F}_{pf}^\nu=&n_p\tilde{u}^\nu_p\frac{g_\pi}{2\pi^2}\int_{m_\pi}^\infty\d \tilde{p}_f^0~ \sqrt{(\tilde{p}_f^0)^2-m_N^2} \\
    &\times\frac{\sqrt{(s-m_N^2-m_\pi^2)^2-4m_N^2m_\pi^2}}{2[\exp(\tilde{p}^0/T_f)-1]}~\sigma_{N\pi\to X}(s)\nonumber
\end{align}
and again analogously for $R_{B,p}$ without the factor of $\tilde{u}_p^\nu$ and dividing by $m_N$. The remaining integral has to be computed numerically. In the form of Eq.~\eqref{eq:fpfric_Lorentz}, it is obvious that this integral is a Lorentz scalar, so the transformation back to the CMS frame of the full collision system simply amounts to replacing $\tilde{u}_p^\nu$ by ${u}_p^\nu$.

\section{Unification procedure}\label{app:unification}

By construction, the splitting of the system into multiple fluids is meaningful if they are clearly separated in momentum space. However, when fluids are close in momentum space, the splitting induces an artificial increase in degrees of freedom, such that it would not describe the correct equilibrium limit. To avoid this, the fluids have to be unified when their flow velocities get close to each other. In our implementation, we locally unify by feeding energy and momentum from projectile and target cells to the fireball cell at the same point in the same way that friction works.  We interpolate between the usual friction terms and unification friction terms based on the ratio of the thermal velocity of projectile or target to their respective relative velocity to the fireball. The unification friction corresponds to an exponential decay of the contents of the projectile or target cell on a unification timescale $\tau_{\rm uni}$, which is a model parameter.  In the calculations presented here, we used $\tau_{\rm uni}=0.2$ fm.

More specifically, the charge and energy-momentum transfer from the projectile to the fireball is given as 
\begin{align}
    R=&(1-\lambda)R_{\rm fric}+\lambda R_{\rm uni}\;,\\
    F_{}^\nu=&(1-\lambda)F_{\rm fric}^\nu+\lambda F_{\rm uni}^\nu\;,\\
    \lambda=& \begin{cases} 
      0 \;, & v_{pf}> v_{T,p}, \\
      \exp\Big[\big(1-{v_{T,p}^2}/{v_{pf}^2}\big)^{-1}\Big]\;,&v_{pf}\leq v_{T,p},\;
   \end{cases} 
\end{align}
where $v_{pf}$ is the relative velocity of the projectile and fireball and $v_{T,p}$ is the thermal velocity of the projectile, computed as the velocity corresponding to the approximate mean Lorentz $\gamma$-factor of a Fermi-distribution in its restframe.
\begin{align}
    v_{T}=&\sqrt{1-\gamma_{T}^{-2}}\;,\\
    \gamma_{T}=&\left\{\int\frac{\d^3p}{(2\pi)^3}\big[z^{-1}\exp(p^0/T)+1\big]^{-1}\right\}^{-1}\\
    &\times\int\frac{\d^3p}{(2\pi)^3}\frac{p^0}{m}\big[z^{-1}\exp(p^0/T)+1\big]^{-1}\nonumber\\
    =&\frac{\mathrm{Li}_2(-ze^{-1/x})+3x\mathrm{Li_3(-ze^{-1/x})+3x^2\mathrm{Li}_4(-ze^{-1/x})}}{\mathrm{Li}_2(-ze^{-1/x})+x\mathrm{Li}_3(-ze^{-1/x})}\\
    \approx&\frac{1}{1+x}+3x\;,
\end{align}
where $z=e^{\mu/T}$ and $x=T/m$.

For a timestep of size $\Delta\tau$, the unification friction for the projectile is given as
\begin{align}
    \Delta\tau R_{\rm uni}&=N_p^\tau[1-\exp(\Delta\tau/\tau_{\rm uni})]\;,\\
    \Delta\tau F_{\rm uni}^\nu&=T^{\tau\nu}_p[1-\exp(\Delta\tau/\tau_{\rm uni})]\;.
\end{align}

\end{document}